%% file: main.tex
\documentclass[10pt,twocolumn,letterpaper]{article}

\usepackage[pagenumbers]{iccv} % To force page numbers, e.g. for an arXiv version

\usepackage[accsupp]{axessibility}

\definecolor{iccvblue}{rgb}{0.21,0.49,0.74}
\usepackage[pagebackref,breaklinks,colorlinks,allcolors=iccvblue]{hyperref}

\input{math_commands.tex}

\usepackage{url}
\usepackage[capitalize]{cleveref}
\usepackage{graphicx}
\usepackage{booktabs}
\usepackage{multirow}
\usepackage{color}
\usepackage{algorithm}
\usepackage{algpseudocode}
\usepackage{comment}

\def\paperID{0000}
\def\confName{ICCV}
\def\confYear{2025}

\title{Touching Space: Accessible Map Exploration Through Conversational Audio-Haptic Interaction}

\author{
Li Liu$^{1}$ \qquad
Jiaming Qu$^{2}$\thanks{Work does not relate to the author's position at Amazon.} \qquad
Marc Jowell Bagaoisan$^{3}$ \qquad
David T. Lee$^{1}$ \qquad
Leilani H. Gilpin$^{1}$\thanks{Corresponding author: lgilpin@ucsc.edu}\\
$^{1}$University of California, Santa Cruz \qquad
$^{2}$Amazon \qquad
$^{3}$University of California, Los Angeles\\
{\tt\small \{lliu112,dlee105,lgilpin\}@ucsc.edu,qjiaming@amazon.com,mjbagaoisan@ucla.edu}
}

\begin{document}
\maketitle

\input{tex/abstract}

\begin{figure*}[t]
  \centering
  \includegraphics[width=\textwidth]{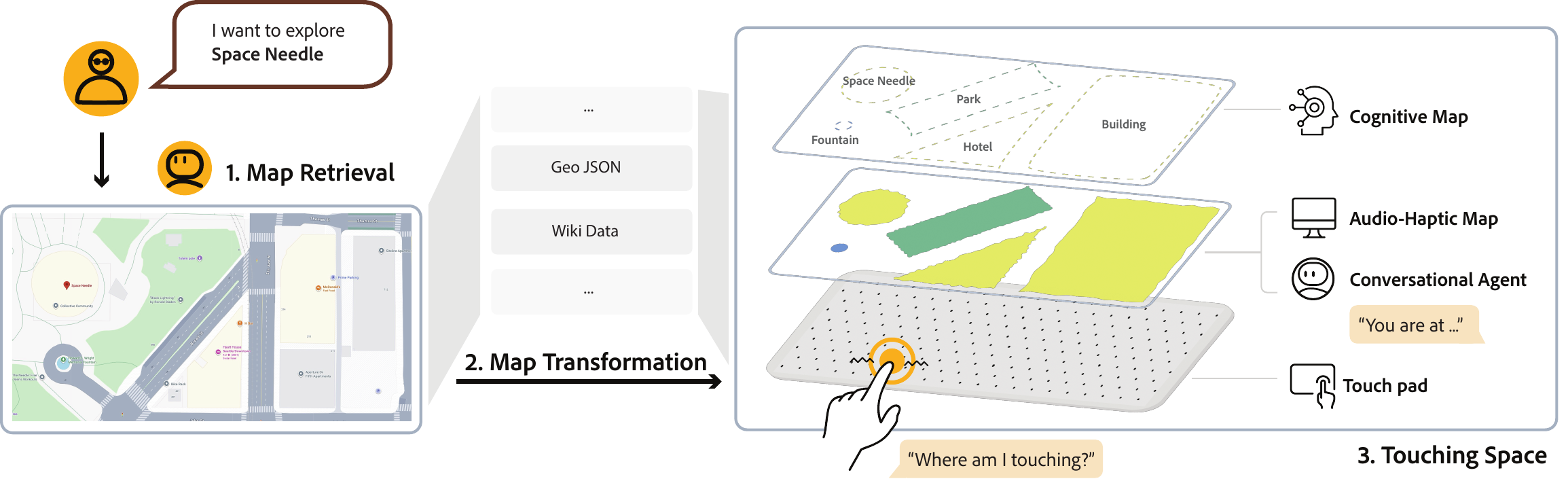}
  \caption{Touching Space: an audio-haptic map system with conversational agent support for spatial learning. (1) The system retrieves raw map data based on requested location. (2) The system transforms geographic data into an audio-haptic map. (3) Users explore the map via touch while the conversational agent answers spatial queries to support cognitive map formation.}
  \label{fig:teaser}
\end{figure*}

\input{tex/introduction}
\input{tex/related_work}

\input{tex/system}

\input{tex/scenario}

\input{tex/discussion}

% \clearpage
% \input{tex/appendix}

{
    \small
    \bibliographystyle{ieeenat_fullname}
    \bibliography{reference}
}

\end{document}

%% file: math_commands.tex
\usepackage{amsmath,amsfonts,bm}

\def\eqref#1{equation~\ref{#1}}
\def\1{\bm{1}}

\DeclareMathAlphabet{\mathsfit}{\encodingdefault}{\sfdefault}{m}{sl}
\SetMathAlphabet{\mathsfit}{bold}{\encodingdefault}{\sfdefault}{bx}{n}

%% file: tex/abstract.tex
\begin{abstract}
Most existing assistive navigation tools focus on providing real-time guidance for Blind and Low-Vision (BLV) people, but few support building a holistic spatial understanding of unfamiliar environments before travel. Such \emph{cognitive map} construction (e.g., knowing that a fountain is south of a tower and west of a hotel) is important for pre-travel planning, yet remains underexplored in prior work. To address this gap, we present \emph{Touching Space}, an end-to-end system that retrieves map data for a target place and loads it into a frontend interface for exploration. The system combines haptic and audio feedback: users explore spatial layouts through touch and ask spoken questions to a conversational agent during exploration. \emph{Touching Space} contributes a conversational interface that supports BLV users in building cognitive maps on commodity hardware.
\end{abstract}

%% file: tex/introduction.tex
\section{Introduction}

Digital navigation technologies have substantially improved mobility access for Blind and Low-Vision (BLV) people. However, most assistive navigation tools focus on real-time, turn-by-turn guidance~\cite{chen2015blindnavi,pohovey2025beyond}. These systems mainly support route knowledge (procedural information about how to move) but focus less on survey knowledge (a broader understanding of spatial layout and landmark relationships)~\cite{siegel1975development}. Survey knowledge is important for helping BLV people build \emph{cognitive maps}---mental models of space that support reasoning about shortcuts, landmark relationships, and orientation when plans change~\cite{tolman1948cognitive,soltani2025user}. However, designing systems for cognitive map construction before travel remains challenging because BLV users differ in vision conditions, navigation strategies, and spatial experiences~\cite{williams2013pray}.

Prior work has explored several approaches to supporting BLV users' spatial understanding. Physical tactile maps and 3D-printed models can convey spatial layout easily, but they are costly, static, and difficult to use on demand~\cite{touchmapper,shi2020molder,nagassa20233d}. Audio interfaces are more accessible, but they present information sequentially, which makes it harder to understand an overall layout~\cite{gonzalez2025tapnav,kuriakose2020multimodal}. Recent AI-powered systems can generate richer descriptions of places~\cite{jain2025scenescout}, but the interaction is still largely one-way: users receive information, but cannot easily ask follow-up questions during exploration. As a result, these approaches do not fully support active exploration, which prior work suggests benefits spatial learning~\cite{chrastil2012active, qin2024active}.

In this work, we present \emph{Touching Space}, a system that combines two complementary modalities to support BLV users in building cognitive maps: (1) continuous vibrotactile feedback on a standard trackpad and (2) a multimodal large language model (LLM)-based conversational agent that users can interact with during exploration. The system has two main parts: a backend pipeline for map retrieval and preprocessing, and a frontend interface that supports audio-haptic exploration. As users move their finger across the trackpad to explore the spatial structure of a place, they can ask open-ended questions, such as identifying landmarks, describing directions and distances, or planning routes between locations. To support these questions, we develop a conversational agent, and it generates responses from a multimodal prompt that integrates four sources of information. Through this prototype, we show how grounding conversational interaction in haptic exploration can support BLV users' cognitive map construction for unfamiliar environments on commodity hardware.

%% file: tex/related_work.tex
\section{Related Work}

\textbf{Cognitive Maps and Non-Visual Spatial Learning.}
Cognitive maps are people's internal spatial representations that support flexible reasoning about environments~\cite{tolman1948cognitive}. Prior work shows that cognitive maps develop through active engagement with space~\cite{siegel1975development}, and that active exploration leads to better spatial knowledge than passive observation~\cite{qin2024active,brunec2023exploration}. For BLV people, studies have shown that cognitive maps can be built through non-visual modalities given appropriate tools~\cite{giudice2018navigating, tivadar2023learning}, and that multimodal input can support richer spatial representations than a single modality~\cite{ottink2022cognitive_review}. However, most assistive navigation systems still focus on turn-by-turn guidance rather than broader spatial understanding~\cite{pohovey2025beyond}. Pre-travel approaches based on auditory overviews~\cite{aziz2022planning}, AI-generated scene descriptions~\cite{jain2025scenescout}, and conversational street-view exploration~\cite{froehlich2025making} improve access to unfamiliar environments, but they mainly make visual content verbally accessible. In contrast, our work supports cognitive map construction by grounding spatial dialogue in haptic exploration.

\textbf{Haptic and Multimodal Spatial Interfaces.}
Studies have shown that interactive audio-tactile maps, in which users touch map elements and receive audio feedback, improve spatial recall and usability compared to static tactile maps~\cite{brock2015interactivity,griffin2020effectiveness,sargsyan2024audio}. Multi-sensory systems that combine haptic and audio feedback can also support cognitive map construction~\cite{lahav2008construction,ottink2022cognitive_review}. However, a recent review shows that haptic tools for BLV users mainly support spatial perception tasks such as shape exploration and navigation, while semantic and relational context remains limited~\cite{jiang2025can}. Recent work has started to address this gap. For example, Jiang et al.~\cite{jiang2025llm_electrotactile} combined an LLM assistant with electrotactile feedback for map previewing. Our work extends this line of research by supporting open-ended questions about semantic and relational information in the user's current exploration context.

\textbf{Conversational Interfaces for Spatial Accessibility.}
Conversational AI is increasingly used in accessibility contexts. Prior work has shown that LLM-based systems can improve accessibility workflows for blind users~\cite{kodandaram2024enabling}, support exploration of unfamiliar shopping malls through open-ended questions~\cite{kaniwa2024chitchatguide}, and augment tactile maps with dialogue and voice question answering~\cite{manzoni2025mapio,cavazos2019jido}. Prior work in cognitive science further suggests that interactive dialogue can support deeper learning than passive information delivery~\cite{chi2001learning,graesser2014learning}, especially when dialogue is anchored to objects both parties can perceive and reference~\cite{clark1991grounding}. Our system extends these in two ways: (1) users can explore maps for any location of interest, and (2) the conversational agent generates responses based on rich interaction data containing cursor position, map screenshot, spatial metadata, and chat history.

%% file: tex/system.tex
\section{Touching Space System}

\emph{Touching Space} is a macOS system that supports BLV users' pre-travel cognitive map construction for unfamiliar environments. The system consists of a backend pipeline for map retrieval and preprocessing (Section~\ref{subsec:backend}) and a frontend interface for audio-haptic exploration (Section~\ref{subsec:frontend}). A key component is a conversational agent that users can interact with during exploration (Section~\ref{subsec:conv_agent}), e.g.,  by asking about spatial relationships between buildings or requesting routes between locations.

\subsection{Map Retrieval and Preprocessing}\label{subsec:backend}

Users begin by specifying a location of interest. The system accepts both spoken and text queries and retrieves geographic data through the Overpass API\footnote{\url{https://wiki.openstreetmap.org/wiki/Overpass_API}}.  We set the retrieved map to cover a 400-meter radius of the target location, which is approximately a five-minute walking distance for most people. We use OSMnx~\cite{boeing2017osmnx} to process raw geographic data, transforming polygon geometries (e.g., buildings and parks) into touchable haptic zones, while excluding point-only and linear features such as trails. The processed polygons are stored as a structured dataset with coordinates and metadata. Finally, the polygons are overlaid on an Apple Maps tile layer based on longitude and latitude coordinates (Fig.~\ref{fig:system_demo}). Both layers (i.e., the Apple Maps base tiles and the processed polygons) are captured in screenshots and used as visual prompts for the conversational agent.

\subsection{Interface for Audio-Haptic Exploration}\label{subsec:frontend}
The frontend interface is implemented as a native macOS application in Swift using SwiftUI and MapKit. It loads the processed map data into an interactive interface for exploration, projects stored polygons onto a pixel canvas\footnote{Pixel distances are mapped to real-world meters with uniform scale.}, and renders the zones as overlays on Apple Maps tiles. The interface provides two modalities of feedback.

\textbf{Haptic feedback.} Users explore the map by moving a finger across the trackpad\footnote{Requires MacBook or Magic Trackpad for haptic feedback.}. As the finger crosses different zones, the system delivers vibrotactile feedback through Apple's Core Haptics framework. Different zone types produce distinct vibration patterns varying in frequency, such as rapid pulses for buildings, slower pulses for parks, and softer patterns for water. Empty regions remain silent. If the finger drifts outside the map boundary, a sharp haptic burst signals the edge, followed by spoken guidance such as ``move left to return to the map zone.'' Re-entering the map triggers a confirmation vibration and audio message.

\textbf{Passive audio feedback.} When the cursor enters a meaningful region (i.e., a polygon zone), the system automatically reads the zone name. As shown in Fig.~\ref{fig:system_demo}, the system reads out (1) ``Museum of Pop Culture,'' (2) ``Hyatt House,'' and (3) ``Space Needle'' in sequence as the cursor moves along the red path. This function is enabled by default, but users can disable it to reduce interruption during free exploration. Together, haptic and passive audio feedback help users build spatial mental models more effectively.

\begin{figure*}[htbp!]
    \centering
    \includegraphics[width=\linewidth]{image/demo_system_seattle.png}
    \caption{The frontend interface of \emph{Touching Space} showing a three-turn conversation during map exploration. The map panel sits at the left and the conversation panel shows on the right. The user's finger trajectory (the red path) moves from (1) MoPOP to (2) Hyatt House, and then to (3) Space Needle with different vibration patterns. At each location, the user asks a spatial question and receives a voice response. In this example, response latency is under two seconds on average.}
        % \Description{A screenshot of the Touching Space macOS application. The left panel shows a map of the Seattle Center area with colored polygon overlays representing buildings and parks on top of Apple Maps tiles. The numbered path (1→2→3) traces the user's finger trajectory through three numbered locations: Museum of Pop Culture, Hyatt House, and Space Needle. A yellow star marks the current cursor position at the Space Needle. The right panel shows a conversation thread with three question-answer pairs, where the user asks spatial questions and receives text responses with timestamps.}

    \label{fig:system_demo}
    \vspace{-.3cm}
\end{figure*}

\subsection{Conversational Agent}\label{subsec:conv_agent}
Our main contribution is a conversational agent that answers users' open-ended spatial questions in addition to reading out predefined metadata for locations. We draw on three design principles from prior work on BLV people's spatial learning: (1) spatial learning benefits from active exploration~\cite{chrastil2012active, qin2024active}, (2) cognitive maps are built progressively~\cite{siegel1975development}, and (3) spatial understanding requires both geometric and semantic information~\cite{giudice2018navigating}. Based on these principles, we designed the agent to prioritize user-initiated dialogue and to ground responses in interaction context.

\textbf{User-initiated dialogue.} Drawing on cognitive load theory~\cite{sweller1988cognitive}, prior work suggests that simultaneous audio and haptic feedback can increase cognitive demands during spatial exploration for BLV people~\cite{kuriakose2020multimodal}. We therefore separate passive audio feedback from on-demand conversation. By default, the system automatically reads out zone names when the cursor enters a region. This provides lightweight identification and can be turned off at any time. For richer spatial inquiry, users can force-click the trackpad to begin a voice query. This on-demand design helps preserve haptic exploration as the primary channel for understanding spatial structure.

\textbf{Grounding responses in interaction context.} Prior research has shown that effective communication relies on a collaborative process in which people continuously establish and update mutual understanding using all available evidence~\cite{clark1991grounding}. We therefore designed a multimodal prompt structure for the conversational agent rather than relying only on the current cursor position. Each time the user asks a question, the system constructs a prompt with information from four sources (Table \ref{tab:prompt_schema}). First, it captures a screenshot of the current map view, including zone overlays, place labels, and the cursor marker. Second, it generates structured spatial context by listing nearby zones together with their compass directions and real-world distances from the current cursor position. For example, the prompt may include a sentence such as ``Sather Tower is to your northeast, about 60 meters away.'' Third, it includes a sliding window of recent conversation history and visited locations to support follow-up questions across turns. Fourth, it appends the user's current question. Motivated by the finding that BLV users often rely on body-centered reference frames when reasoning about space~\cite{thinus1997representation}, the agent responds in egocentric terms, using the cursor as the reference point, such as ``to your northeast'' or ``move your finger to the upper-right.''

% \begin{table}[t]
% \caption{Multimodal prompt schema sent to the LLM. Each query combines visual, spatial, and conversational context grounded in the user's current cursor position.}
% \label{tab:prompt_schema}
% \small
% \centering
% \begin{tabular}{@{}p{0.12\linewidth} p{0.15\linewidth} p{0.08\linewidth} p{0.52\linewidth}@{}}
% \toprule
% \textbf{Category} & \textbf{Component} & \textbf{Modality} & \textbf{Content} \\
% \midrule
% Static & System instruction & Text & Role prompt: ``You are a guide helping a blind user explore a haptic map\ldots'' with response rules (egocentric language, concise answers). \\
% \midrule
% \multirow{2}{*}{Chat history} 
%   & Chat log & Text & Up to 20 prior turns.\\
% & Visited locations & Text & Chronological list: \texttt{(1.Space Needle, 2.McCaw Hall)} \\
% \midrule
% \multirow{3}{*}[-1.5em]{\shortstack[l]{Spatial\\Context}} 
%   & Map screenshot & Image & JPEG screenshot of current canvas, including zone overlays on Apple Maps tiles and a $\bigstar$ cursor marker at the user's current position. \\
%   & Current zone & Text & \texttt{CURRENT ZONE: McCaw Hall (building)} \\
%   & Spatial layout & Text & Nearest 8 zones with direction and distance:\newline \texttt{- MoPOP (building) is to your SE about 170\,m away}\newline \texttt{- Fountain Lawn (park) is to your W about 180\,m away} \\
% \midrule
% User query & User question & Text & ``Am I still west of the park?'' \\
% \bottomrule
% \end{tabular}
% \end{table}

\begin{table*}[!htp]
\centering
\resizebox{0.9\textwidth}{!}{%
\begin{tabular}{@{}p{0.12\linewidth} p{0.15\linewidth} p{0.08\linewidth} p{0.52\linewidth}@{}}
\toprule
\textbf{Category} & \textbf{Component} & \textbf{Modality} & \textbf{Content} \\
\midrule
Static & System instruction & Text & Role prompt: \textit{``You are a guide helping a blind user explore a haptic map\ldots''} with response rules (egocentric language, concise answers). \\
\midrule
\multirow{2}{*}{Chat history} 
  & Chat log & Text & Up to 20 prior turns.\\
& Visited locations & Text & Chronological list: \textit{(1.Space Needle, 2.McCaw Hall)} \\
\midrule
\multirow{3}{*}[-1.5em]{\shortstack[l]{Spatial\\Context}} 
  & Map screenshot & Image & JPEG screenshot of current canvas, including zone overlays on Apple Maps tiles and a $\bigstar$ cursor marker at the user's current position. \\
  & Current zone & Text & Current zone: \textit{Space Needle (building)} \\
  & Spatial layout & Text & Nearest 10 zones with direction and distance:\newline - \textit{MoPOP (building) is to your SE about 170\,m away\newline - Fountain Lawn (park) is to your W about 180\,m away} \\
\midrule
User query & User question & Text & \textit{``Am I still west of the park?''} \\
\bottomrule
\end{tabular}
}
\caption{Multimodal prompt schema sent to the LLM. Each query combines visual, spatial, and conversational context grounded in the user's current cursor position.}
\label{tab:prompt_schema}
\end{table*}

We experimented with two implementation paths for the conversational agent in the current prototype. The first is a \emph{voice-streaming} pipeline, where raw audio is sent directly to a multimodal speech API (Gemini Live\footnote{\url{https://ai.google.dev/gemini-api/docs/live}}) for spoken responses. The second is a \emph{text-mediated} pipeline, where the system first transcribes user speech on-device using Apple's Speech framework, then queries a vision-language model (Gemini Flash\footnote{\url{https://deepmind.google/models/gemini/flash/}} or Qwen3-VL~\cite{bai2025qwen3}), and finally synthesizes speech through Apple TTS or the Gemini TTS API. Both pipelines receive the same prompt structure: a map screenshot, structured spatial metadata encoding compass directions and distances, and a sliding window of up to 20 conversation turns. Our preliminary evaluation found that the text-mediated pipeline produces more stable latency and more structured answers, but its speech output sounds more mechanical; in contrast, the voice-streaming pipeline sounds more natural but has more variable latency depending on network conditions.

%% file: tex/scenario.tex
\section{Interaction Walkthrough}
\label{scenario}

% \clearpage
\begin{table*}[t]
\centering
\resizebox{0.9\textwidth}{!}{%
\begin{tabular}{@{}p{0.2\linewidth} p{0.27\linewidth} p{0.42\linewidth}@{}}
\toprule
\textbf{Query Type} & \textbf{Example} & \textbf{Purpose} \\
\midrule
Identification & ``What am I touching?'' & Connects haptic region to place name. \\
Spatial relationship & ``What is around this building?'' & Describes nearby zones with directions. \\
Comparison & ``Which building is bigger?'' & Supports relative judgments about size or distance. \\
Guidance & ``Where is the library from here?'' & Provides trackpad-relative directional guidance. \\
Contextual knowledge & ``What is this building used for?'' & Provides additional information about places. \\
Confirmation & ``Is the museum still to my east?'' & Verifies or corrects spatial understanding. \\
\bottomrule
\end{tabular}
}
\caption{Example query categories during map exploration. The agent accepts open-ended natural-language questions. Note these categories illustrate common interaction patterns rather than an exhaustive list.}
\label{tab:qa_types}
\end{table*}

We illustrate how users interact with the system using the Seattle Center area as an example (Fig.~\ref{fig:system_demo}).

A user begins by sweeping their finger across the trackpad. As the cursor crosses different zones, the system provides distinct vibration patterns and reads out zone names. These cues help the user detect nearby regions, but touch alone does not fully convey each zone's feature or their relation to surrounding landmarks. When the passive audio reads out ``Museum of Pop Culture,'' the user force-clicks and asks: \textbf{``Hello, where am I?''} The agent responds with a short description of the current location, helping the user identify the first landmark on the map with richer context.

The user then continues exploring and arrives at a zone labeled ``Hyatt House.'' To understand how it relates to the previous landmark, the user asks: \textbf{``How can I get to the Space Needle?''} The agent responds: \textit{``The Space Needle is to your northwest, about 100 meters away. Move your finger to the northwest to reach it.''} The user follows this guidance, briefly crossing an open area before reaching the next zone. After arriving there, the user asks: \textbf{``Where is Pacific Science Center?''} In this way, the user incrementally builds a cognitive map by alternating between haptic exploration and spoken questions.

These examples illustrate a common interaction pattern in \emph{Touching Space}: users first locate regions through touch and passive audio, then ask open-ended questions to identify landmarks, understand spatial relationships, and guide further exploration. Table~\ref{tab:qa_types} shows example query categories. Through these queries, the conversational agent extends the other two interaction channels. For \emph{touch}, conversation helps interpret vibrotactile patterns by answering questions about shape, identity, and nearby landmarks. For \emph{audio labels}, it turns a spoken place name into richer and more personalized information. More broadly, this active process lets users iteratively explore, ask, verify, and refine their understanding of the space, supporting progressive cognitive map construction.

%% file: tex/discussion.tex
\section{Discussion and Conclusion}
\textbf{Summary.} In this paper, we presented \emph{Touching Space}, a system that supports BLV users' cognitive map construction for unfamiliar places before travel. The system takes a target location as input, retrieves and processes map data, and loads it into a frontend interface where users can explore with haptic and audio feedback. Haptic feedback conveys spatial structure through touch, helping users perceive shapes, boundaries, and spatial density. Passive audio feedback reads out zone names for immediate identification. More importantly, \emph{Touching Space} introduces a conversational agent that answers open-ended spatial questions. The agent generates responses based on a multimodal prompt that integrates four sources of information, which enables it to support landmark identification, spatial reasoning, and exploration guidance during map exploration. Together, these components show how audio-haptic interaction and conversational interaction can jointly support cognitive map construction on commodity hardware.

\textbf{Limitations and Future Work.} 
This work takes an initial step toward conversational support for BLV users' pre-travel spatial learning. The current system has several limitations. First, the conversational agent can still produce hallucinated responses (e.g., incorrect directions or distances) despite the multimodal prompt design. We aim to develop better grounding strategies that cross-check spatial claims against verified sources. Second, the current prototype relies on off-the-shelf LLMs and does not yet incorporate external retrieval or other fact-checking mechanisms, which may be especially important for handling more complex spatial questions. Third, we have not yet evaluated the system with real BLV users. A key next step is therefore a user study comparing audio-haptic map exploration with and without conversational support, examining spatial knowledge gain, workload, and user strategies during cognitive map construction. More broadly, future work should also investigate whether knowledge built through such pre-travel interaction transfers to real-world wayfinding.